\documentclass[allclo]{FBSart}
\usepackage{amsfonts}
\usepackage{amssymb}

\def\frac#1,#2{{#1\over #2}}

\def\fr#1,#2{{#1\over #2}}

\def\ha{\frac{1}{2}}

\def\lan{\langle}
\def\ran{\rangle}

\def\bra#1{\lan#1|}
\def\ket#1{|#1\ran}
\def\ha{{1\over 2}}

\def\normord#1{\mathopen{\hbox{\bf:}}#1\mathclose{\hbox{\bf:}}}
\def\be{\begin{equation}}
\def\ee{\end{equation}}
\def\bea{\begin{eqnarray}}
\def\eea{\end{eqnarray}}

\title{Induced Operators in QCD\footnote{Presented at Light-Cone 2004, Amserdam, 16 - 20 August}}
\author{G. McCartor}
\institute{Department of Physics, SMU, Dallas, TX 75275, USA}

\runningauthor{G. \ McCartor}
\runningtitle{Induced Operators in QCD}
\begin{document}

\maketitle

\enlargethispage*{50pt}

\begin{abstract}
Light-cone quantization always involves the solution of differential
 constraint equations.  The solutions to these equations include 
 integration constants (fields independent of $x_-$).  These fields 
 are unphysical but when they are consistently removed from the 
 dynamics, additional operators (induced operators), which would not 
 be present if the integration constants were simply set to zero, are 
 included in the dynamics.  These induced operators can be taken to 
 act in the usual light-cone subspace, for instance, the space used 
 for DLCQ.  Here, I shall give a derivation of two such operators.  
 The operators are derived starting from the QCD Lagrangian but the 
 derivation involves some guesses.  The operators will provide for the 
 linear growth of the pion mass squared with the quark bare mass and for the 
 splitting of the pi and the rho at zero quark mass.
\end{abstract}

\section{Introduction}

I shall discuss some work that I have done with Simon Dalley \cite{dm} 
on induced operators in QCD.  I shall begin with a very brief review of the 
Schwinger model, the one model that includes induced operators that can 
be analyzed in complete detail \cite{mc99}.  The reasons for reviewing the Schwinger model are to remind ourselves that induced operators do exist and do have physical consequences, and to set some notation and ideas for the discussion of QCD, where the analysis cannot be done in complete detail.  I shall then derive two of the induced operators in QCD and illustrate aspects of their effects with simple models.  I shall end with a brief discussion.

\section{Review of the Schwinger model}

When we solve the Schwinger model in the light-cone representation \cite{nm00}we must solve the light-cone constraint equation
\be
   \partial_- \psi_- = - {\rm i} \mu \psi_+ \; ,
\ee
as
\be
    \psi_- = \psi_-^0(x^+) - {\rm i}\ha\mu \int \psi_+ dx^- \; ,
\ee
where
\be
   \psi_-^0 =  Z_-(\mu){\rm e}^{\Lambda_-^{(-)}(\mu)}\sigma_- {\rm e}^{\Lambda_-^{(+)}(\mu)} \; .
\ee
Here, $\sigma_-$ is a spurion and $\Lambda_-$ is the fusion field.
The spurion, $\sigma_-$, is needed to construct the physical vacuum as:
\be
\ket{\Omega(\theta)}  \equiv  \sum_{M=-\infty}^\infty {\rm e}^{{\rm i}M\theta}\ket{\Omega(M)} \quad;\quad  \ket{\Omega(M)} = ({\sigma}_+^*{\sigma}_-)^M \ket{0} \; .
\ee
The field $\Lambda_-$ is entirely unphysical, but if we write the physical field $\psi_+$ in bosonized form as
\be
      \psi_+ = Z_+ {\rm e}^{\Lambda_+^{(-)}}\sigma_+ {\rm e}^{\Lambda_+^{(+)}} \; ,
\ee
we find that the operator $\bar{\psi} \psi$ includes the term \cite{mc99}
\be
\mu \int \normord{\bar{\psi} \psi} \ dx^-  \supset   \mu Z_-Z_+  \int_{-\infty}^\infty \left(\sigma_-^*\sigma_+ ({\rm e}^{\Lambda_+^{(-)}}{\rm e}^{\Lambda_+^{(+)}} - 1)  + C.C.\right) dx^- \; . \label{io}
\ee
Here, we have not included the fields $\Lambda_-$ since they do not act in the physical subspace.  Also, $\sigma_-^*\sigma_+$ acts as a constant in the physical subspace so the physical subspace can be taken to be the usual light-cone subspace created by operators from the $\psi_+$ field but with the addition of the operator (\ref{io}) with $\sigma_-^*\sigma_+$ replaced by the appropriate c-number.  The operator (\ref{io}) is an induced operator.  It is important to understand that all the above results can be obtained by quantizing the model either on the light-cone or at equal-time, so the conclusions are independent of quantization surface.  Details, including the operator solution can be found in \cite{nm00}.

Before leaving the Schwinger model there is one further point we should discuss since it will be important to what follows.  If we consider the case where periodicity conditions are imposed on $x^+ = 0$ (DLCQ) then most of the results obtained in the continuum still follow.  But the coefficient of the induced operator goes to zero as the periodicity length, $L$, goes to infinity.  That is because the induced operator is proportional to the chiral condensate which in the continuum is given by
\be
    \bra{\Omega(\theta)} {\bar{\psi} \psi}_1 \ket{\Omega(\theta)} = - \frac{e},{2 \pi^{3/2}} e^\gamma \cos{\theta} \; ,
\ee
but in the periodic case is given by \cite{mc94}
\be
\bra{\Omega(\theta)} \bar{\psi} \psi \ket{\Omega(\theta)} = - \fr{1},{L} \cos{\theta} \; .
\ee
The fix for the problem is to multiply the induced operator in the periodic case by a renormalization constant which is equivalent to taking
\be
     Z_- =  \fr{e L {\rm e}^\gamma},{2\pi^{3/2}} \; .
\ee
With this correction, the all the results in the periodic case go to those in the continuum case when $L \rightarrow \infty$.  We do not know any way to determine the correct coefficient in the periodic case other than to compare with the continuum.  If we had access to two dimensional experimentalists that could be comparing with experiment.

\section{QCD}

We now turn our attention to QCD using the Schwinger model as a guide \cite{dm,mc00}.  We quantize the fermi fields on $x^+ = 0$ in the standard way:
\be
 \psi^{(a)}_{+,s}(0,x^-,x^\perp)  =  {1\over\sqrt{\Omega}}\sum_{k_\perp}{\rm e}^{ik_\perp x^\perp} \tilde{\psi}^{(a)}_{+,s}(0,x^-,k^\perp) \; ,
\ee
where 
\be
\tilde{\psi}^{(a)}_{+,s}(0,x^-,k^\perp)   = 
\sum_{n = 1}^\infty 
   b^{(a)}_s(n,-k_\perp) {\rm e}^{-ik_-(n)x^-} +
   d^{(a)*}_{-s}(n,k_\perp,) {\rm e}^{ik_-(n)x^-} \; . \label{ff}
\ee
The object in (\ref{ff}) is a standard one dimensional fermi field and so, can be bosonized in the standard way.  We write
\be
\tilde{\psi}^{(a)}_{+,s}(0,x^-,k^\perp) = {\rm e}^{-\lambda_s^{(a)(-)}(x^-,k_\perp)}
\sigma_{+,s}^{(a)}(x^-,k_\perp)
{\rm e}^{-\lambda_s^{(a)(+)}(x^-,k_\perp)} \; ,
\ee
where
\be
\lambda_s^{(a)(+)}(x^-,k_\perp) = -\sum_{n=1}^\infty
 {1 \over n}C^{(a)}_s(n,k_\perp){\rm e}^{-i\tilde{k}_{-}(n)x^-} \; ,
\ee
\be
\lambda_s^{(a)(-)}(x^-,k_\perp) = -{\lambda_s^{(a)(+)}}^* =  \sum_{n=1}^\infty
 {1 \over n}C^{(a)*}_s(n,k_\perp){\rm e}^{i\tilde{k}_{-}(n)x^-} \; ,
\ee
and
\bea
C^{(a)}_s(n,k_\perp)  &=&   \sum^{n-1}_{\ell = 0} d^{(a)}_{-s}\left(\ell + \frac{1},{2},k_\perp\right) b^{(a)}_s\left(n
- \ell - \frac{1},{2},-k_\perp\right) +  \nonumber \\
&&  \sum^{\infty}_{\ell = 0} b^{(a)*}_s \left(\ell + 
\frac{1},{2},-k_\perp\right) b^{(a)}_s
\left(\ell + n + \frac{1},{2},-k_\perp\right) -  \nonumber \\ 
&&   \sum^{\infty}_{\ell = 0}d^{(a)*}_{-s} \left(\ell + \frac{1},{2},k_\perp\right) d^{(a)}_{-s}
\left(\ell + n + \frac{1},{2},k_\perp\right)
\eea

In QCD, the field $\psi_-$ satisfies a constraint equation and the solution to that equation involves an integration constant, $\psi^0_-$, (a function of $x^+$ and $x^\perp$).  We specify that integration constant as 
\begin{equation}
   \psi^{0(a)}_-(x^+,x^\perp) = 
{1\over\sqrt{\Omega}}\sum_{s,k_\perp}{\rm e}^{ik_\perp x^\perp}\sum_{n = 1}^\infty 
   \beta^{(a)}_s(n,-k_\perp) {\rm e}^{-ik_+(n) x^+} +
   \delta^{(a)*}_{-s}(n,k_\perp) {\rm e}^{ik_+(n) x^+} 
\end{equation}
As with the case of the Schwinger model, the only physical operator from this field is the spurion, $\sigma_{-,s}^{(a)}$ (the other operators in $\psi^0_-$ are needed to maintain the canonical structure of the theory).

We shall need only the transverse components of the gluon fields.  We work in the helicity basis defined by
\be
  A_{\uparrow} = (A_1 -{\rm i}A_2)/\sqrt{2} \;,\; A_{\downarrow} = (A_1 +{\rm i}A_2)/\sqrt{2} \; ,
\ee
then quantize as
\be
 A^{(c)}_s(0,x^-,x^\perp)  =  
{1\over\sqrt{\Omega}}\sum_{k_\perp}{\rm e}^{ik_\perp x^\perp} 
\tilde{A}^{(c)}_s(0,x^-,k^\perp) \; ,
\ee
where
\bea 
   &&\tilde{A}^{(c)}_s(0,x^-,k^\perp)     =  \nonumber \\
 &&\sum_{n = 1}^\infty 
   \frac{1},{\sqrt{2 \tilde{k}_-(n)}}\left(a^{(c)}_s(n,-k_\perp) \, {\rm e}^{-i\tilde{k}_-(n)x^-} +
   {a^{(c)}_{s}}^*(n,k_\perp,)  \,  {\rm e}^{i\tilde{k}_-(n)x^-} \right) \; .
\eea

So far everything has just been algebra.  Now we need an ansatz for the vacuum and we shall use the Schwinger model for a guide.  The Schwinger model might suggest that we require
\be
        \sigma_{-,-s}^{(a)*}(x^-,k_\perp)\sigma_{+,s}^{(a)}(x^-,k_\perp) \ket{\Omega} = \kappa \ket{\Omega} \; . \label{vev}
\ee
The vector which satisfies that requirement is
\be
\ket{\Omega} = \prod_{s\, ; \, a\, ; \, k_\perp}\left(\sum_{n =-\infty}^\infty (\kappa \, \sigma_{-,-s}^{(a)*}(0,k_\perp) \, \sigma_{+,s}^{(a)}(0,k_\perp))^n \right) \ket{0} \; .
\ee
That vector is not a suitable choice for the vacuum since it is not gauge invariant.  But we find that if we choose the gauge invariant state
\be
\ket{\Omega} = \prod_{s\, ; \, k_\perp}\left(\sum_{n =-\infty}^\infty (\kappa \sum_{a} \, \sigma_{-,-s}^{(a)*}(0,k_\perp) \, \sigma_{+,s}^{(a)}(0,k_\perp))^n \right) \ket{0} \label{vac} \; ,
\ee
to be the vacuum we can work out the action of the induced operators in the physical subspace assuming that (\ref{vev}) holds even though (\ref{vac}) does not satisfy that relation in the entire space.

With all this machinery in place we can specify the induced operators.  The first one, which we call $I_1$, is a part of the term $\bar{\psi} \psi$.  If we keep only the part of $\bar{\psi} \psi$ that depends on $\psi^0_-$ (actually, on $\sigma_{-,s}^{(a)}$).  We have
\be
{I}_1  =  
\mu  \int dx^-\;d^2x^\perp \left({\psi^{(a)0}_{-,\downarrow}}^* \psi^{(a)}_{+,\uparrow} + {\psi^{(a)}_{+,\uparrow}}^* \psi^{(a)0}_{-,\downarrow}  + {\psi^{(a)0}_{-,\uparrow}}^* \psi^{(a)}_{+,\downarrow} + {\psi^{(a)}_{+,\downarrow}}^* \psi^{(a)0}_{-,\uparrow}\right)
\ee
which, keeping only the spurion from $\psi^0_-$ , can be rewritten as
\bea
{I}_1  &=&   \mu g_1  \int dx^-\;d^2k^\perp   \Bigl( \sigma_{-,\downarrow}^{(a)*}(0,k_\perp)\sigma_{+,\uparrow}^{(a)}(0,k_\perp)\Bigr)\nonumber \\
&&\left({\rm e}^{-\lambda_\uparrow^{(a)(-)}(x^-,k_\perp)}
{\rm e}^{-\lambda_\uparrow^{(a)(+)}(x^-,k_\perp)} -1 \right) 
  + C.C. + spin flip \Bigr) \; ,
\eea
where $g_1$ is a constant included to account for the fact that, as in the case of the Schwinger model, the operator may scale in a nontrivial way with the periodicity length $L$.  Since we do not have a continuum solution with which to compare, $g_1$ will have to be fit to a symmetry or to data.  Now, we can commute the combination, $\sigma_{-,\downarrow}^{(a)*}(0,k_\perp)\sigma_{+,\uparrow}^{(a)}(0,k_\perp)$, or its adjoint, to the vacuum where, by (\ref{vev}), it will become a c-number, either $\kappa$ or $\kappa^*$ (if we wish to have a C-invariant theory $\kappa$ must be real).  Once we have eliminated $\sigma_{-,\downarrow}^{(a)*}(0,k_\perp)\sigma_{+,\uparrow}^{(a)}(0,k_\perp)$ in favor of $\kappa$, the remaining operators are operators that act in the usual DLCQ subspace and we can perform the rest of the calculation in that space taking the vacuum to be the bare light-cone vacuum.

As an example of the action of $I_1$ I shall give its action on a valence type state for a meson.  We have
\bea
&&I_1 \; {b^{(a)}}^*_\uparrow(n,-k_\perp) {d^{(a)}}^*_\downarrow(m,k_\perp) \ket{0} =   
  \mu g_1 \kappa \Bigl({b^{(a)}}^*_\uparrow(n-1,-k_\perp) {d^{(a)}}^*_\downarrow(m + 1,k_\perp) \nonumber \\
 &&- 2  {b^{(a)}}^*_\uparrow(n,-k_\perp) {d^{(a)}}^*_\downarrow(m,k_\perp)  +  {b^{(a)}}^*_\uparrow(n + 1,-k_\perp) {d^{(a)}}^*_\downarrow(m - 1,k_\perp)\Bigr) \ket{0} \; .
\eea
We find that if we restrict ourselves to meson valence states of the form
\be
\int_0^1 dx \; f(x)\, {b^{(a)}}^*_\uparrow(x)\, {d^{(a)}}^*_\downarrow(1-x) \; ,
\ee
then:  if $g_1$ is independent of $L$, $I_1$ goes to zero as $L$ goes to infinity; if $g_1$ scales linearly with $L$, $I_1$ becomes a pure endpoint operator as $L$ goes to infinity; if $g_1$ is proportional to $L^2$, $I_1$ becomes the operator $\partial_x$ as $L$ goes to infinity.  If $I_1$ does not go to zero, it will give the relation that the mass of the pion squared is proportional to the bare quark mass for small bare quark mass.

Now let us develop the other induced operator, $I_2$, the induced operator that is part of the term $J_\perp A^\perp$.  We write
\bea
 &&I_2   =    I_{2,1} + I_{2,2} + I_{2,3} + I_{2,4} = g \int dx^-\;d^2x^\perp \sum_{abc}\lambda_{a b}^c \nonumber \\
 &&\left( \psi_{+,\downarrow}^{(a)*}  \psi_{-,\downarrow}^{0 (b)} A_\uparrow^{(c)} -\psi_{-,\uparrow}^{0(a)*} \psi_{+,\uparrow}^{(b)} A_\uparrow^{(c)} + \psi_{-,\downarrow}^{0(a)*}  \psi_{+,\downarrow}^{(b)} A_\downarrow^{(c)}  - \psi_{+,\uparrow}^{(a)*}  \psi_{-,\uparrow}^{0(b)} A_\downarrow^{(c)} \right)
\eea
where $g$ is the usual QCD coupling constant.  Consider the first of these operators
\bea
I_{2,1} &=& g g_2  \int d^2p_\perp d^2k_\perp dx^- \sum_{abc} \lambda_{a b}^c \nonumber \\
   &&\tilde{\psi}_{+,\downarrow}^{(a)*}(x^-,k_\perp)  
\sigma_{-,\downarrow}^{(b)}(x^-,k_\perp - p_\perp)
\tilde{A}_\uparrow^{(c)}(x^-,p_\perp)
\eea
where, as was the case with $g_1$, $g_2$ is a constant incorporating the unknown dependence of the coefficient of $I_2$ on the periodicity length $L$.  We now insert the number 1 written as $\sigma_{+,\uparrow}^{(b)*}(x^-,k_\perp - p_\perp)
\sigma_{+,\uparrow}^{(b)}(x^-,k_\perp - p_\perp)$ to get
\bea
I_{2,1} = -g g_2  \int d^2p_\perp d^2k_\perp dx^- \sum_{abc}&&\lambda_{a b}^c\tilde{A}_\uparrow^{(c)}(x^-,p_\perp)
   \tilde{\psi}_{+,\downarrow}^{(a)*}(x^-,k_\perp) 
   \sigma_{+,\uparrow}^{(b)}(x^-,k_\perp - p_\perp) \cr 
&&\Big[\sigma_{-,\downarrow}^{(b)}(x^-,k_\perp - p_\perp)\sigma_{+,\uparrow}^{(b)*}(x^-,k_\perp - p_\perp)\Big]
\eea
We can commute the operator in brackets through any other operators until it accts on the vacuum at which time it will just give a c-number.  After performing that operation all the remaining operators will act in the usual DLCQ subspace and we an perform the remainder of the calculations in that space taking the bare light-cone vacuum as the vacuum.  Working similarly, we find that the rest of $I_2$ is given by
\bea
I_{2,2} = g g_2 \int d^2p_\perp d^2k_\perp dx^- \sum_{abc} && \lambda_{a b}^c\tilde{A}_\uparrow^{(c)}(x^-,p_\perp)
{\sigma}_{+,\downarrow}^{(a)*}(x^-,k_\perp+ p_\perp)  \tilde{\psi}_{+,\uparrow}^{(b)}(x^-,k_\perp )  \nonumber \\
&&\Big[\sigma_{-,\uparrow}^{(a)*}(x^-,k_\perp + p_\perp)\sigma_{+,\downarrow}^{(a)}(x^-,k_\perp + p_\perp)\Big]
\eea
\bea
I_{2,3} = - g g_2  \int d^2p_\perp d^2k_\perp dx^-  \sum_{abc} &&\lambda_{a b}^c\tilde{A}_\downarrow^{(c)}(x^-,p_\perp)
{\sigma}_{+,\uparrow}^{(a)*}(x^-,k_\perp + p_\perp)  \tilde{\psi}_{+,\downarrow}^{(b)}(x^-,k_\perp )  \nonumber \\
&&\Big[\sigma_{-,\downarrow}^{(a)*}(x^-,k_\perp + p_\perp)\sigma_{+,\uparrow}^{(a)}(x^-,k_\perp + p_\perp)\Big]
\eea
\bea
I_{2,4} = g g_2  \int d^2p_\perp d^2k_\perp dx^- \sum_{abc} &&\lambda_{a b}^c \tilde{A}_\downarrow^{(c)}(x^-,p_\perp)
\tilde{\psi}_{+,\uparrow}^{(a)*}(x^-,k_\perp)  \sigma_{+,\downarrow}^{(b)}(x^-,k_\perp - p_\perp)  \nonumber \\
&&\Big[\sigma_{-,\uparrow}^{(b)}(x^-,k_\perp - p_\perp)\sigma_{+,\downarrow}^{(b)*}(x^-,k_\perp - p_\perp)\Big]
\eea
To illustrate the action of $I_2$ we work out its action on a valence type meson state.  We find that if we define $\ket{s} = b^{(d)*}_\downarrow(n,-q_\perp) d^{(d)*}_\uparrow(K-n,q_\perp) \ket{\Omega}$ then
\bea
&&I_2 \; \; \ket{s} =  \left( I_{2,1} + I_{2,3} \right) \; \; \ket{s} \; ,
\eea
where
\bea
&&I_{2,1} \ket{s} = - g g_2 \kappa^* \int d^2p_\perp \;  \sqrt{\frac{\pi},{K-n-1}}\sum_{abc}\lambda^c_{ab} \nonumber \\
&&b^{(a)*}_\downarrow(n,-q_\perp)d^{(b)*}_\downarrow(1,q_\perp - p_\perp) a^{(c)*}_\uparrow(K-n-1,p_\perp)\ket{0} \; , 
\eea
and
\bea
&&I_{2,3} \ket{s} =- g g_2 \kappa \int d^2p_\perp \;  \sqrt{\frac{\pi},{n-1}}\sum_{abc}\lambda^c_{ab} \nonumber \\
&&b^{(b)*}_\uparrow(1,-q_\perp - p_\perp)d^{(a)*}_\uparrow(K-n,q_\perp) a^{(c)*}_\downarrow(n-1,p_\perp) \ket{0}
\eea
The pattern seen here is general: although the form of $I_2$ may appear to be complicated, its action is easy to state.  It acts successively on each quark just as a standard QCD three point vertex except that the quark spin is always flipped and the gluon absorbs all the available longitudinal momentum leaving the quark in the lowest longitudinal momentum state.  Color works as in the standard vertices and the transverse momentum is shared between the quark and the gluon in all possible ways just as in the standard vertices.  The operator is nonzero at zero quark bare mass and will split the pi and rho masses at zero bare quark mass.

Let us further illustrate the action of $I_1$ and $I_2$ by solving a simple model.  We shall keep only $I_1$ and $I_2$ and the quark kinetic energy as our $P^-$.  We shall take the quark bare mass to be $\mu$, $\kappa$ to be -1 and $g_1$ to be $g_1 = g_{11} K^2$ where $K$ is the DLCQ harmonic resolution.  We can find an eigenstate for the pion that has the form
\bea
\ket{\pi} = 
 &&C \left(a_\downarrow^*(1)b^*_\uparrow(0)\, d^*_\uparrow(0) +  
a_\uparrow^*(1) b^*_\downarrow(0)\, d^*_\downarrow(0)\right)\ket{0} \nonumber \\
+ && \int_0^1 dx \;\pi(x)\, {1\over \sqrt{2}}\left(b^*_\uparrow(x)\, d^*_\downarrow(1-x) - b^*_\downarrow(1-x)\, d^*_\uparrow(x)\right)\ket{0} \; .
\eea
With the wave function in that form, $I_1$ has the form $I_1 \rightarrow - \mu g_1 \fr{\partial^2},{\partial x^2}$ in the large $K$ limit.  Thus, in that limit, $\pi(x)$ satisfies the equation
\be
- \mu g_{11} \frac{d^2\pi(x)},{dx^2} + \frac{\mu^2},{x(1-x)} \pi(x) = M^2 \pi(x) \; ,
\ee
where $M$ is the mass of the pion.  To get a finite, nontrivial result we must choose $g_2$ to have the form
\be
g_2 = \fr{2},{g} \, \sqrt{ g_{11} \, \mu^3} \,  K^{\frac{3},{2}} - \fr{8 \mu^3},{3 g g_{11}} \sqrt{2 K} \ln{K} - g_{22} \sqrt{K} \; ,
\ee
where $g_{22}$ is a free parameter.  With that choice we find that
\be
M^2_\pi = \frac{8 \mu^3},{g_{11}} + 4 g g_{22} \sqrt{\frac{\mu},{g_{11}}} \; ,
\ee
and
\bea
\pi(x) &=&  \sqrt{\frac{\mu},{g_{11}}} +\left(\frac{\mu},{g_{11}}\right)^{\frac{3},{2}} x (1-x)\ln{[x(1-x)]} + \nonumber \\
&&\frac{2 g g_{22}},{g_{11}^2} x (1-x) + \left(\frac{\mu},{g_{11}}\right)^{\frac{3},{2}} x^2 (1-x)^2\ln{[x(1-x)]} \; .
\eea
We can find an eigenstate of the rho of the form
\be
\ket{\rho} =  
  \int_0^1 dx \;\rho(x)\, {1\over \sqrt{2}}\left(b^*_\uparrow(x)\, d^*_\downarrow(1-x) + b^*_\downarrow(1-x)\, d^*_\uparrow(x)\right)\ket{0} \; .
\ee
For values of $\mu$ that are not too large we find that, to a good approximation
\be
\rho(x) \approx \sqrt{2} \sin{\pi x} \quad ; \quad M^2_\rho = \mu \pi^2 g_{11} + 4.88 \mu^2 \; .
\ee
As an example, if we choose $\mu = 5$, $g_{11} = 1.25 \times 10^4$ and $g g_{22} = 2.45 \times 10^5$ we get $g g_{21} = 2500$,  $M_\rho = 785$ and $M_\pi = 140$.  While, given the mass dimensions of the various constants, these numbers are all consistent with $\Lambda_{QCC}$ being a few hundred MEV, the model is not very realistic in that $I_2$ not only splits the pi and rho masses but gives the rho all of its mass (except for a small contribution from the kinetic energy).  Presumably the rho gets most of its mass from other operators in the $P^-$ of QCD that are left out of this simple model.

Further samples of model calculations with different scalings of the coefficients of the induced operators may be found in ref. \cite{dm}, including a DLCQ calculation of dimensionally reduced gauge theory that includes the standard part of $P^-$ as well as the induced operator $I_2$.  In all the cases studied the pi and rho masses are split in the continuum limit.

\section{Summary}

We have given two of the induced operators in QCD.  These arise due to the fact that there are (unphysical) quarks in the vacuum state.  There will probably be other induced operators due to the presence of (unphysical) gluons in the vacuum; we do not yet have a form for these operators or for the glue in the vacuum.  It is possible that the induced operators we have given here are the most important ones for determining the structure of low mass Hadrons but we do not know that.

To derive the induced operators we made an ansatz for the vacuum based on the Schwinger model.  In principal it is possible to check whether or not that ansatz is correct.  But to do so we would need a nonperturbative regulator for the theory that preserves gauge invariance and Lorentz invariance.  We hope some day to have such a regulator but we do not yet have one that we can use to check the ansatz for the vacuum.  As they stand, the induced operators have coefficients that may scale with the DLCQ harmonic resolution and whose values are unknown.  In principal these values, including the scaling, are determined by requiring that all the fields are canonical at all space-like separations in the continuum case.  We are unable to carry out the analysis required to use that principle to determine the coefficients and must fit them either to a symmetry or to data.

We have used the induced operators in calculations of simple models in two dimensions.  We hope soon to extend these studies to four dimensions and to include the effects of additional flavors.  On the basis of the results from the simple models it is clear that the operators can split the masses of the pion and the rho at small values of the quark mass, can give a linear growth of the mass squared of the pion with bare quark mass and can modify the endpoint behavior of the pion distribution function.

\begin{acknowledge}
This work was supported in part by the U.S. Department of Energy under contract DE-FG03-95ER40908.
\end{acknowledge}

\end{document}